# Implementing the Independent Reaction Time method in Geant4 for radiation chemistry simulations

Mathieu Karamitros[1*], Jeremy M. C. Brown[2], Nathanael Lampe[3], Dousatsu Sakata[4], Ngoc Hoang Tran[3], Wook-Guen Shin[3], Jose Ramos Mendez[5], Susana Guatelli[4], Sébastien Incerti[3], Jay A. LaVerne[1]

(1) Radiation Laboratory, University of Notre Dame, Notre Dame, In 46556, U.S.A.
(2) Radiation Science and Technology, Delft University of Technology, Delft, The Netherlands
(3) Université de Bordeaux, CNRS, IN2P3, CENBG, UMR 5797, 333170 Gradignan, France
(4) Centre For Medical Radiation Physics, University of Wollongong, Wollongong, Australia
(5) Department of Radiation Oncology, University of California San Francisco, San Francisco CA, 94115, USA.

* mat.kara@cern.ch

## Abstract

The Independent Reaction Time method is a computationally efficient Monte-Carlo based approach to simulate the evolution of initially heterogeneously distributed reaction-diffusion systems that has seen wide-scale implementation in the field of radiation chemistry modeling. The method gains its efficiency by preventing multiple calculations steps before a reaction can take place. In this work we outline the development and implementation of this method in the Geant4 toolkit to model ionizing radiation induced chemical species in liquid water. The accuracy and validity of these developed chemical models in Geant4 is verified against analytical solutions of well stirred bimolecular systems confined in a fully reflective box.

## 1. Introduction

In liquids, such as water, ionizing radiation produces isolated clusters of reactants, called spurs, that may contain only a small number of reactants (2-6 species). The initial distribution of spurs can also be highly heterogeneous. In such regimes, deterministic descriptions of kinetics are usually not well suited, and a spatial stochastic description of the diffusion-reaction system is needed. Developed in the 1980's to satisfy these constraints, the Independent Reaction Times (IRT) method[1–5] is nowadays widely used in radiation chemistry[6–10]. This method uses a particle-based (or particle-continuum) representation of the chemical system where all reactants of interest are modelled as spheres, whereas the solvent is modelled as a continuum.

The IRT method consists in breaking down the n-body problem to a two-body problem. The reaction time of a pair of reactants is sampled by inverting a known probability function depending on the initial separation distance of the pair and few parameters, such as the diffusion coefficients and the reaction rate constant. The pair of reactants having the smallest reaction time is selected as the next reaction. Compared to the step-by-step approach described in a previous paper[11], this method has the advantage of producing a reaction at each time step.

The models used in this method are based on the backward Smoluchowski diffusion equation. Both the diffusion-controlled reaction and partially diffusion-controlled reaction models are derived from setting boundary conditions on the solution of the backward Smoluchowski differential equation to account for the reaction at encounter with or without intrinsic reaction probability. In this work we outline our implementation of the IRT method, based on these solutions to the backward Smoluchowski differential equation, in the Monte Carlo radiation toolkit Geant4[12–14] through its low energy extension for biological applications Geant4-DNA[15–18].

## 2. Theoretical background of the IRT method

The following provides a summary of the IRT method to explain our work in the context of its adaption and implementation in the Monte Carlo radiation transport modelling toolkit Geant4. We consider a pair of reactants A and B in solution, where B is diffusing in the reference frame centered on A. The forward Debye Smoluchowski (FDS) equation describes the kinetics of species B in the solution:

$$\partial_t p = \vec{\nabla} D (\vec{\nabla} p + \beta \cdot p \cdot \vec{\nabla} U)$$

where $p \equiv p_B$ is the density distribution of the species B, $D$ is the sum of the diffusion coefficients of A and B, $\beta = \frac{1}{k_B T}$ where $k_B$ is the Boltzmann constant, $T$ is the system's temperature, and $U(r)$ is a central potential energy between A and B.

The fundamental solution of the FDS equation is the Green function $p(\vec{r}, t | \vec{r_0})$ defined as $p(\vec{r}, 0 | \vec{r_0}) = \delta(\vec{r} - \vec{r_0}) = \frac{\delta(r - r_0) \delta(\theta - \theta_0) \delta(\phi - \phi_0)}{r^2 \sin(\theta)}$, where $p(\vec{r}, t | \vec{r_0})$ has units of $[length]^{-3}$. As we are interested in the separation distance between the species, then:

$$p(r, t | r_0) = \frac{1}{4\pi} \int_0^{2\pi} d\phi \int_0^{\pi} d\theta \, \sin \theta \, p(\vec{r}, t | \vec{r_0})$$

where $\theta$ is the polar angle and $\phi$ is the azimuthal angle matching the ISO convention for spherical coordinates. From this we can construct the density probability function[a] $4\pi r^2 \cdot p(r,t|r_0)$ that describes the probability of finding reactant B in $[r, r+dr]$ based on the initial radial separation $r_0$ of the pair at initial time, and time elapsed $t$. Substitution of this into the FDS equation then reduces to[11]:

$$\frac{\partial p}{\partial t}(r,t|r_0) = \frac{1}{r^2}\partial_r\left(r^2 \cdot j(r,t|r_0)\right)$$

with

$$j(r,t|r_0) = D \cdot e^{-\beta U(r)}\left[\partial_r\left(e^{\beta U(r)} \cdot p(r,t|r_0)\right)\right].$$

With this form of the FDS, we apply the following boundary conditions[11]:

$$j(r,t|R) = w \cdot p(r,t|R)$$

$$p(r,t_0|r_0) = \frac{\delta(r-r_0)}{4\pi r^2}$$

where $R$ is a critical radius in which absorption may take place, also called reaction radius, $\delta(x)$ is the Dirac function, and $w$ is the so-called reaction velocity (has units of $[length][time]^{-1}$). Here $p(r,t_0|r_0)$ has units of $[length]^{-1}$, and $j(r,t|R)$ has units of $[time]^{-1}$. However, when calculating the pair survival probability, it is easier to manipulate the backward Debye-Smoluchowski (BDS) equation:

$$\partial_{t_0} p(\vec{r},t_f|\vec{r_0},t_0) = D\left(\Delta_0 p(\vec{r},t_f|\vec{r_0},t_0) - \beta \cdot \nabla_0 U \cdot \nabla_0 p(\vec{r},t_f|\vec{r_0},t_0)\right).$$

In the BDS equation, the operator $\Delta_0$ is the Laplace operator applied on the initial position $\vec{r_0}$. After integrating over both the angles in spherical space, the radial BDS equation reduces to:

$$\partial_t p(r,t|r_0) = D\left(\frac{1}{r_0}\partial_0^2(r_0 \cdot p(r,t|r_0)) - \beta \cdot \partial_0 U \cdot \partial_0 p(r,t|r_0)\right)$$

where $t$ is simply the elapsed time (i.e. $t = t_f - t_0$), and $\partial_0$ is the differential operator with respect to $r_0$. In contrast to the FDS case, it can be shown that boundary conditions of the BDS are independent of the external field:

---

[a] This distribution is intentionally divided by $\frac{1}{4\pi r^2}$ so that the forward Debye-Smoluchowski equation is applicable to $p(r,t|r_0)$. This definition leads to the initial condition $p(r,t=0|r_0) = \frac{\delta(r-r_0)}{4\pi r^2}$. Otherwise, one can define $p(r,t|r_0) = \int_0^{2\pi} d\phi \int_0^{\pi} d\theta\, r^2 \cdot \sin\theta\, p(\vec{r},t|\vec{r_0})$ to yield $p(r,t=0|r_0) = \delta(r-r_0)$. Then $p(r,t|r_0)$ describes the density probability function of finding the reactant B in $[r, r+dr]$.

$$D \cdot \partial_{r_0} p \Big|_{r_0 = R} = w \cdot p(R, t | r_0)$$

based on the initial condition that:

$$p(r, t = 0 | r_0) = \frac{\delta(r - r_0)}{4\pi r^2} \text{ with } r_0 > R.$$

Note that in this situation, the $r_0$ belongs to $[R, +\infty]$ where $R$ is the reaction radius, and that $\int_R^\infty \delta(r - r_0) \cdot dr_0 = 1$.

For the reactant pair A and B, their survival probability $S(t|r_0, R)$ can defined as:

$$S(t|r_0, R) = \int_R^\infty 4\pi r^2 \cdot p(r, t|r_0) \cdot dr$$

and the pair reaction probability $W(t|r_0, R)$, the survival probabilities counterpart, is given by:

$$W(t|r_0, R) = 1 - S(t|r_0, R).$$

Here, both the pair survival and the pair reaction probabilities follow the BDS equation with the following set of initial and boundary conditions:

| | |
|---|---|
| $S(0\|r_0, R) = 1$ with $r_0 > R$ | $W(0\|r_0, R) = 0$ with $r_0 > R$ |
| $S(t\|r_0 \to \infty, R) = 1$ with t finite | $W(t\|r_0 \to \infty, R) = 0$ with t finite |
| $S(t \to \infty\|r_0, R) = S_\infty(r_0)$ | $W(t \to \infty\|r_0, R) = W_\infty(r_0)$ |
| $S(t\|r_0 = R, R) = 1 - P_{react}$ | $W(t\|r_0 = R, R) = P_{react}$ |
| $D \cdot \partial_{r_0} S\big|_{r_0=R} = w \cdot S(t\|r_0, R)$ | $D \cdot \partial_{r_0} W\big|_{r_0=R} = w \cdot (W(t\|r_0, R) - 1)$ |

where $P_{react}$ is the reaction probability at encounter and it equals to unity when considering diffusion-controlled reactions. $S_\infty(r_0)$ can also be written as $S_\infty(r_0) = 1 - \Pr[t_r < \infty]$ where $t_r$ is the reaction time and $W_\infty(r_0) = \Pr[t_r < \infty]$ is the probability that the pair will ultimately react (viz. the reaction will happen at finite time).

The approach of the IRT method consists in determining the $W(r, t|r_0)$ function for a given reaction type. $W(r, t|r_0)$ can either be determined analytically or numerically. Alternatively, in some extreme cases, asymptotic approximations can be obtained. Knowing the initial separation distance of a reactant pair, the $W(r, t|r_0)$ function is inverted to sample a random reaction time. If the inverse of $W(t|r_0, R)$ with respect to $t$ is analytically unknown, numerical techniques can be applied to inverse the reaction probability. Several root finding algorithms exist, such as Newton's method which is simple to apply and can deal with arbitrary forms of continuously differentiable functions.

We should note here that at every time step, only one reaction or a subset of all possible reactions is selected. Therefore, in principle, the new positions of the non-immediately reacting species should be sampled with the condition that no reaction happened during the corresponding time step.

### 1. Reaction between neutral species

If $U(r) = 0$, then the solution to the Green function representing the Smoluchowski differential equation can be written[19]

$$4\pi r^2 \cdot p(r,t|r_0) = \frac{r}{2r_0\sqrt{\pi Dt}} \left(\exp(-u^2) + \exp(-v^2) - 2x\sqrt{\pi} \cdot Wc(x,v)\right)$$

with[a] $W_C(a,b) = \exp(a^2 + 2ab) \cdot \text{erfc}(a+b)$ and

$$\begin{cases} u = \dfrac{r-R}{2\cdot\sqrt{Dt}}, & B = \dfrac{w}{D} + \dfrac{1}{R} = \dfrac{1}{R}\dfrac{k_{act}}{k_{obs}} \\ v = \dfrac{r+r_0-2R}{2\cdot\sqrt{Dt}}, & x = B\cdot\sqrt{Dt} \end{cases}$$

The time dependent reaction rate $k_{obs}(t)$ is generally described as the composition of two internal rates: the so-called encounter rate $k_{diff}$ describing the encounter probability of reactant pairs, and the so-called activation rate $k_{act}$ describing the reaction rate only once the species have encountered one another. Note that in general, both of these internal rates may be time dependent. $k_{obs}$ is the ultimate observer reaction rate constant corresponding to $k_{obs} = k_{obs}(t \to \infty)$.

As the pair reaction probability $W(t|r_0,R)$ is given by:

$$W(t|r_0,R) = 1 - S(t|r_0,R)$$

then the reaction probability for partially diffusion-controlled reactions between neutral species can be written:

$$W(t|r_0,R) = W_\infty \cdot (\text{erfc}(y) - W_C(x,y)) \quad (1)$$

where we pose that:

$$\begin{cases} W_\infty = \dfrac{1}{r_0}\left(R - \dfrac{1}{B}\right) = \dfrac{R}{r_0}\left(1 - \dfrac{k_{obs}}{k_{act}}\right) \\ y = \dfrac{r_0 - R}{2\cdot\sqrt{Dt}} \end{cases}$$

---

[a] $\text{erfc}(x) = \dfrac{2}{\sqrt{\pi}}\int_x^\infty \exp(-t^2)dt = 1 - \text{erf}(x)$

From this, when B is finite and $t \to \infty$ or $t \to 0$, then $x + y \to \infty$ and $W_C \to 0$. Alternatively, when $t$ is finite and $B \to \infty$, $x + y \to \infty$ and $W_C \to 0$. In both cases, we find the diffusion-controlled reaction model case for neutral species defined when $k_{act} \to \infty$ that can be written:

$$W(t|r_0, R) = W_\infty \cdot \text{erfc}(y) \tag{2}$$

with

$$W_\infty = \frac{R}{r_0}.$$

Finally, the effect of the partially diffusion-controlled reaction model is significant when $r_0 \to R$, corresponding to a dense environment, and that for a given $t$ the value of $W_C$ approaches an upper limit (i.e. $W_C \sim \frac{k_{obs}}{k_{act}} \frac{R}{\sqrt{Dt\pi}}$).

### 2. Reaction between ionic species

For ionic reactions, Clifford, Green and their coworkers[1,20–23] derived approximations to the probability of isolated ionic pair reactions for both diffusion-controlled reactions:

$$W(t|r_0, R) = W_\infty(r_0, R) \cdot \text{erfc}(y) \tag{3}$$

and partially diffusion-controlled reactions:

$$W(t|r_0, R) = W_\infty(r_0, R) \cdot (\text{erfc}(y) - W_C(x, y)) \tag{4}$$

that are valid for an Onsager radius $r_C = k_C \frac{q_1 q_2}{\epsilon k_B T}$ ($|r_c| \leq 10$ Å or $|r_c| \sim R$), or $\sim 7$ Å in the case of liquid water at room temperature. Here:

$$\begin{cases} x = \frac{\alpha R^2}{D r_c^2} \sqrt{16Dt} \sinh^2\left(\frac{r_C}{2R}\right) \\ y = r_C \cdot \frac{\coth\left(\frac{r_C}{2r_0}\right) - \coth\left(\frac{r_C}{2R}\right)}{\sqrt{16Dt}} \\ \alpha = w + \frac{r_C D}{R^2 \left(1 - \exp\left(-\frac{r_C}{R}\right)\right)} \end{cases}$$

with $q_1$ and $q_1$ giving the charges of two ions, $k_C$ the Coulomb constant and $\epsilon$ being the relative permittivity of the medium. It should be noted that $r_C$ can be positive or negative depending on the charges of the reactants. Furthermore $W_\infty$, the probability of the final reaction (i.e. $t \to \infty$), can be expressed for partially diffusion-controlled reactions as:

$$W_\infty(r_0, R) = \frac{1 - \exp\left(\frac{r_C}{r_0}\right)}{1 - \exp\left(\frac{r_C}{R}\right)\left(1 + \frac{D r_C}{w R^2}\right)}$$

and for diffusion-controlled reactions ($w \to \infty$) as:

$$W_\infty(r_0, R) = \frac{1 - \exp\left(\frac{r_c}{r_0}\right)}{1 - \exp\left(\frac{r_c}{R}\right)}$$

when $r_0 \to R$, then $y \to 0$, and $W_c$ approaches an upper limit.

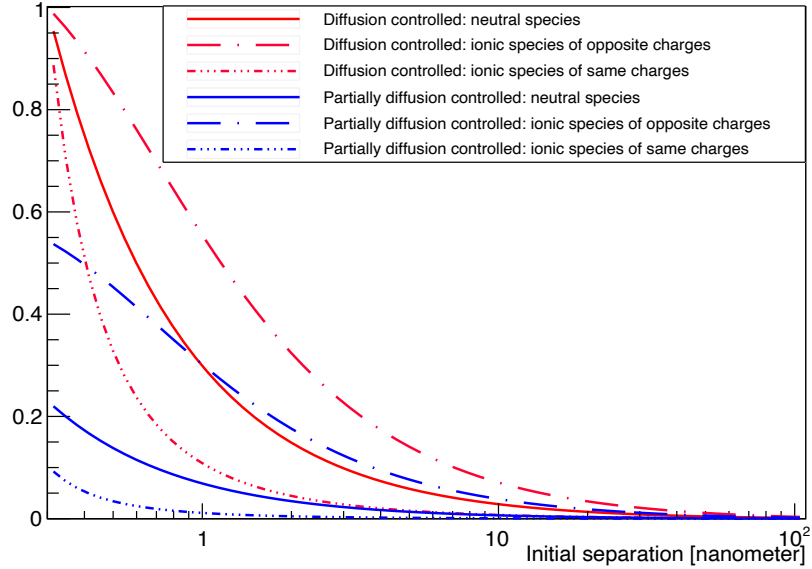

Figure 1 - Reaction probability with respect to the initial separation distance of a reactant pair. Parameters: $R = 0.3\ nm$, $D = 10 \cdot 10^{-9}\ m^2 \cdot s^{-1}, w = 10\ m \cdot s^{-1}, r_c = \pm 0.7\ nm, t = 1\ \mu s$.

An illustration of the reaction probability for six different cases as a function of initial separation distance of the reactant pair can be seen in Figure 1. For ions of opposite charges, the reaction probability is larger than that which can be observed for neutral reactions. Whereas in the case for reactions between ions of the same charge, the reaction probability is decreased in comparison to neutral cases.

It is worth noting that any time reactants where placed such as $r_0 < R$ we considered $y = 0$.

## 3. Relation with the reaction rate constant

Reaction rate constants can be either calculated from an isolated pair or from a homogeneously distribution of reactants. In the following we illustrate the relationship between the model parameters (essentially $R$ and $w$) and the reaction rate constants measured experimentally by considering homogeneously distributed reactants.

### 1. Neutral reactions

The reaction rate constant for a single reactant A surrounded with a homogeneous distribution of reactants B can be defined, on a microscopic scale, as:

$$k_{obs}(t) = 4\pi R^2 \int_R^\infty j(r,t|R) \cdot dr.$$

This reaction rate has units of $[\text{time}]^{-1}$, and represents the number of species $B$ entering a sphere of radius $R$ centered on a single reactant A. As $j(t,r|R) = w \cdot p(r,t|R)$, we can modify:

$$k_{obs}(t) = \underbrace{4\pi R^2 w}_{k_{act}} \cdot S(t|R)$$

through the use of the pair survival relationship above to yield:

$$k_{obs}(t) = \frac{k_{diff} \cdot k_{act}}{k_{diff} + k_{act}} \left(1 + \frac{k_{act}}{k_{diff}} \exp\left(\frac{Dt}{R^2}\left(1 + \frac{k_{act}}{k_{diff}}\right)^2\right) \cdot \text{erfc}\left(\frac{\sqrt{Dt}}{R} \cdot \left(1 + \frac{k_{act}}{k_{diff}}\right)\right)\right)$$

where $k_{diff} = 4\pi RD$ and $k_{act} = 4\pi R^2 w$.

Furthermore, given that the asymptotic expansion of $\text{erfc}(x)$ for a large real value of x can be written as:

$$\text{erfc}(x) = \frac{\exp(-x^2)}{\sqrt{\pi}} \left(\frac{1}{x} - \frac{1}{2x^3} + \cdots\right)$$

this time-dependent reaction rate can be simplified to:

$$k_{obs}(t) \approx \frac{k_{diff} \cdot k_{act}}{k_{diff} + k_{act}} \left(1 + \frac{k_{act}}{k_{act} + k_{diff}} \cdot \frac{R}{\sqrt{\pi Dt}}\right) \qquad (5)$$

which yields the final reaction rate constant when $t \to \infty$ :

$$k_{obs}(\infty) = \frac{k_{diff} \cdot k_{act}}{k_{diff} + k_{act}} \qquad (6)$$

In the case of diffusion-controlled reactions, we assume that $k_{act} \to \infty$ and obtain the following form for the time-dependent reaction rate:

$$k_{obs}(t) \approx k_{diff}\left(1 + \frac{R}{\sqrt{\pi Dt}}\right).$$

### 2. Ionic reactions

It is possible to evaluate the asymptotic reaction rate constant $k_{obs}(\infty)$ for the bulk reactions where there are central forces between the reactants using the following formula[11] :

$$k_{obs}(\infty) = \frac{4\pi}{\frac{\exp(\beta U(R))}{R^2 w} + \mathcal{R}(R)}$$

where:

$$\mathcal{R}(R) = \frac{1}{D}\int_{R_0}^{\infty}\frac{1}{\tilde{r}^2}\exp(\beta U(\tilde{r}))\,d\tilde{r}.$$

In the case of ionic reactions:

$$\mathcal{R}(R) = \frac{1}{r_C D}\left(\exp\left(\frac{r_C}{R}\right) - 1\right)$$

and:

$$k_{obs}(\infty) = \frac{4\pi r_C D}{\exp\left(\frac{r_C}{R}\right)\left(\frac{r_C D}{R^2 w} + 1\right) - 1} = 4\pi \cdot R_{eff}^{obs} \cdot D$$

where:

$$R_{eff}^{obs} = \frac{r_C}{\exp\left(\frac{r_C}{R}\right)\left(\frac{r_C D}{R^2 w} + 1\right) - 1}.$$

If we assume that the diffusion-controlled reactions correspond to $w \to \infty$, then:

$$k_{obs}(\infty)|_{w\to\infty} = \frac{4\pi r_C D}{\exp\left(\frac{r_C}{R}\right) - 1} \equiv k_{diff}.$$

Analogous to the reaction rate constant formula obtained for neutral reactions, the effective reaction radius can be generally defined for ionic reactions as[a]:

$$R_{eff}^{diff} = \frac{r_C}{\exp\left(\frac{r_C}{R}\right) - 1}.$$

Assuming that the expression of the diffusion rate constant $k_{diff}$ does not change in the general case ($k_{act} \neq 0$) due to Noyes' time independent formula, we may then extract the expression of $k_{act}$:

$$k_{act} = \frac{k_{obs}(\infty)\cdot k_{diff}}{k_{diff} - k_{obs}(\infty)} = 4\pi R^2 w \cdot \exp\left(-\frac{r_C}{R}\right).$$

---

[a] The form of $R_{eff}^{diff}$, assuming $R_{eff}^{diff} < 0$, can be written:

$$R_{eff}^{diff} < 0 \Rightarrow \text{or} \begin{cases} \begin{cases} r_C < 0 \\ \exp\left(\frac{r_C}{R}\right) - 1 > 0 \end{cases} \\ \begin{cases} r_C > 0 \\ \exp\left(\frac{r_C}{R}\right) - 1 < 0 \end{cases} \end{cases}$$

which results in the same contradiction in both cases, i.e. $R_{eff}^{diff}$ is always positive. In comparison to the neutral case for a fixed $w$, $k_{act}$ is increased when the charges of the reactants are opposite and decreases when the charge of the reactants are the same.

Note that Green calculated the time-dependent reaction rate in case of ionic reactions[23] which can be expressed as:

$$k_{obs}(t) = 4\pi \cdot R_{eff}^{obs} \cdot D \cdot \left(1 + \left(\frac{wR^2}{D \cdot R_{eff}^{obs}} - 1\right) \cdot \exp(Q^2 \cdot t) \cdot \text{erfc}(Q\sqrt{t})\right)$$

where:

$$Q = \frac{wR^2 \cdot \left(1 - \exp\left(-\frac{r_C}{R}\right)\right)}{r_C \cdot R_{eff}^{obs} \cdot \sqrt{D}}.$$

Substitution of $Q$ and simplification yields:

$$k_{obs}(t) \approx 4\pi \cdot R_{eff}^{obs} \cdot D \cdot \left(1 + \frac{R_{eff}^{obs} \cdot \left(\frac{r_C D}{R^2 w} + 1\right) \cdot \exp\left(\frac{r_C}{R}\right)}{\sqrt{\pi D t}}\right)$$

that can alternatively be expressed as:

$$k_{obs}(t) \approx \frac{k_{diff} \cdot k_{act}}{k_{diff} + k_{act}} \cdot \left(1 + \frac{R_{eff}^{obs} \cdot \left(\frac{r_C D}{R^2 w} + 1\right) \cdot \exp\left(\frac{r_C}{R}\right)}{\sqrt{\pi D t}}\right)$$

For diffusion-controlled reactions, i.e. where $w \to \infty$, $R_{eff}^{obs} = R_{eff}^{diff}$ and

$$k_{obs}(t) \approx 4\pi \cdot R_{eff}^{diff} \cdot D \cdot \left(1 + \frac{R_{eff}^{diff} \cdot \exp\left(\frac{r_C}{R}\right)}{\sqrt{\pi D t}}\right).$$

Thus, for diffusion-controlled reactions, letting $t \to \infty$ means one can write the bimolecular reaction rate constants for both ionic and neutral reactions as

$$k_{obs}(\infty) \approx k_{diff} = 4\pi \cdot R_{eff}^{diff} \cdot D.$$

### 3. Discussion about the reaction velocity

The "reaction velocity" $w$ has the units of velocity and was set to define a radiation boundary condition. Two extreme cases are usually considered; $w \to \infty$ for diffusion-controlled reactions, and $w \to 0$ for reflective encounter (no reaction). Finite $w$ corresponds to partially diffusion-controlled reactions.

Given the general form of $k_{obs}(\infty)$, we can express $k_{act}$ as

$$k_{act} = 4\pi R^2 w \cdot \exp(-\beta \cdot U(R)).$$

In the hard sphere model where $R$ would be chosen to be larger than the radius of the reactants, we can write

$$k_{act} = P_{react} \cdot Z_{coll}^{R_0} \cdot V$$

where $P_{react}$ would be the reaction probability per encounter, $Z_{coll}^{R}$ the collision frequency of the reactants in the sphere of radius $R_0$, and $V$ the volume of the sphere subtended by the radius $R_0$. In this sphere the reactants may re-encounter one another several times before a reaction happens, and would show ballistic behavior rather than Brownian trajectories.

The collision frequency $Z_{coll}^{R_0}$ can expressed based on collision theory as:

$$Z_{coll}^{R_0} = v_{rel} \cdot \sigma_{AB} \cdot \frac{1}{V}$$

where $v_{rel}$ is the relative velocity of the reactants that is assumed to be constant, $\sigma_{AB}$ is the sum of the effective cross sections of the reactants ($\sigma_{AB} = \pi(r_A + r_B)^2$), and $r_A$ and $r_B$ are the radii of the reactant spheres. From this we can construct the following expressions:

$$k_{act} = \sigma_{AB} \cdot P_{react} \cdot v_{rel} = 4\pi R^2 w \cdot \exp(-\beta \cdot U(R)),$$

$$w = \frac{\sigma_{AB} \cdot P_{react} \cdot v_{rel}}{4\pi R^2 \cdot \exp(-\beta \cdot U(R))}.$$

If we assume $r_A + r_B \sim R$ then:

$$w = \frac{P_{react} \cdot v_{rel}}{4} \exp(\beta \cdot U(R)).$$

Using a different approach Naqvi et al[24] arrived at a similar expression that corresponds to the case $U(R) = 0$ of our expression. The authors also implemented a hard sphere point of view and ballistic modeling in the last step of the reaction. Here we only wish to emphasize that $w$ should already depend on the intrinsic reaction probability. Additionally, the diffusion-controlled reaction model assumes that a reaction happens at every encounter, meaning that $P_{react} = 1$. In this condition, we note that defining $w \to \infty$ for diffusion-controlled reactions is only for mathematical convenience.

## 4. IRT model parameters

The computation of the reaction probabilities needs two parameters, namely the reaction radius $R_0$ and the reaction velocity $w$, that can be deduced by the knowledge of two of the three following reaction rate constants $k_{obs}$, $k_{act}$ and $k_{diff}$.

The ultimate observed reaction rate constant is given by the time-independent Noyes formula:

$$k_{obs}(\infty) = \frac{k_{diff} \cdot k_{act}}{k_{diff} + k_{act}}.$$

For reactions between neutral species:

$$k_{diff} = 4\pi \cdot s \cdot D \cdot R$$

$$k_{act} = 4\pi \cdot R^2 \cdot w$$

where $s$ is the spin statistic factor ($s \leq 1$) for radical-radical reactions that depends on the spin relaxation time of the radicals involved which we ignore in this work. From these relations we can write for reactions between neutral species:

$$R = \frac{k_{diff}}{4\pi \cdot s \cdot D}$$

$$w = \frac{k_{act}}{4\pi R^2}$$

with the reactions between ionic species being described by:

$$k_{diff} = \frac{4\pi \cdot s \cdot D \cdot r_C}{\left[\exp\left(\frac{r_C}{R}\right) - 1\right]}$$

$$k_{act} = 4\pi \cdot R^2 \cdot w \cdot \exp\left(-\frac{r_C}{R}\right)$$

which results in:

$$R = \frac{r_C}{\ln\left(1 + \frac{4\pi \cdot s \cdot D \cdot r_C}{k_{diff}}\right)}$$

$$w = \frac{k_{act}}{4\pi R^2 \cdot \exp\left(-\frac{r_C}{R}\right)}$$

Therefore with these expressions, knowledge of $k_{obs}(\infty)$, $k_{act}$, the Onsager radius[a], and the sum of the diffusion coefficients $D$ is all that is needed to determine all the constants used in this model. Finally, if one is interested in the diffusion-controlled reactions with $s = 1$, the only parameter to compute is $R_0$. This can be deduced by letting $k_{obs} \approx k_{diff}$ and knowing $D$.

## 5. Implementation details

As outlined above, the IRT method consists in breaking down the n-body problem to a two-body problem to enable the event-based simulation of radiation chemistry. This is achieved through the construction of an event table that contains the chemical species positions and "reaction time" for every reactant pair of interest. Within this table the entries are sorted in ascending "reaction time" order and only those which fall within a defined "reaction time" interval are taken into account. The reason for this is outlined below.

---

[a] Note that the Onsager radius only depends on the temperature, relative permittivity and charges of the reactants.

Figure 2 presents the stepping algorithm that is utilized at each time step of the simulation. At each time step, the reactions within the event table is selected and processed in ascending "reaction time" order. After a reaction has occurred, the table is updated through removing any listed reactant pairs corresponding to either one of the two chemical species that participated in this reaction. A reaction may produce new products, which are further taken into account in the event table. The individual species positions recorded in the event table are used to calculate the location of the reactants at the time of the reaction[3]. The center of the new positions of the reactants makes it possible to define the reaction location/site. The created product positions are randomly sample within a sphere of radius $R$ centered at the reaction site. In case of reflecting boundary conditions, the product positions may then be further constrained, for instance, using a positioning algorithm similar to that outline in the section 6.

After distributing the generated chemical products, we trigger the diffusion events for the Brownian particles that may have reached the reaction boundaries of the products. This is done using the following approximation: the maximum range for which a chemical species B that can react with the newly positioned chemical product A is limited to $R_d = 8\sqrt{D \cdot dt_s}$. Here D is the diffusion coefficient of the other chemical species B, and this approximation guarantees a 99% coverage of the diffusion range[11] of B during the time step $dt_s$. One can pick $dt_s$ either as the elapsed virtual time since the beginning of the simulation, or as the maximum elapsed virtual time of the particles of type B since the last position update. This choice does not significantly affect the results. To preserve this event-based behavior, and for our benchmarking investigation outlined in section 6, we developed a diffusion mechanism that accounts for reflections at the boundary of a box without explicitly taking diffusion steps into account.

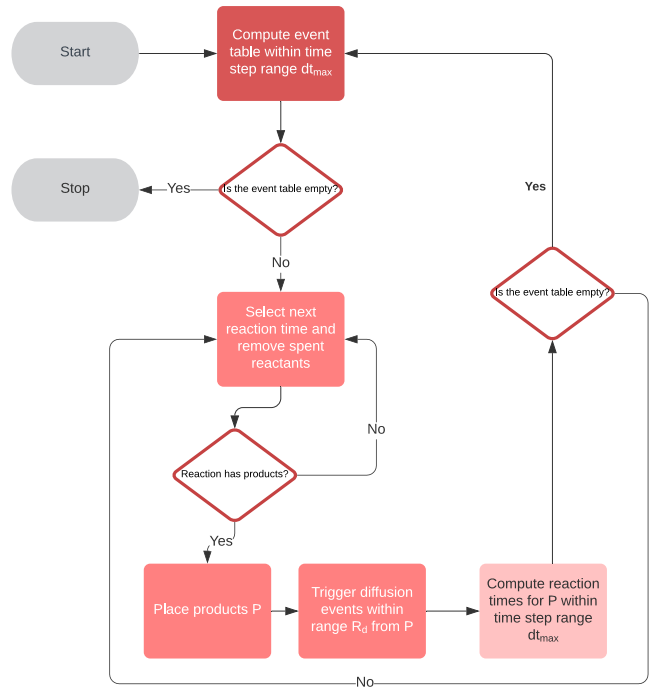

Figure 2 – The computational flow process of the IRT stepping algorithm implemented in Geant4.

## 6. Benchmarking the IRT models

To verify the validity of the implemented IRT models, we compare its results to that of an analytical model of a sufficiently large number of chemical species distributed homogenously within a reflective box for AA and AB reactions. Here, we assume that the initial conditions of the box mimic a well stirred environment, and the reaction rate of the solution can be model with the second order rate law:

$$\frac{d[Z]}{dt} = \sum_x \pm k_{x+y} \cdot [x][y]$$

where reactions $x + y$ are either sources or sinks of a given species Z.

### 1. The analytical reflective box model

The analytical reflective box model is based on an algorithm that samples the next position of a Brownian particle located in a reflective box after a time step $t$. This approach was implemented to prevent the simulation from stopping any time a particle reaches the edge of the box. The Green function solution of this problem, in Cartesian coordinates, can be written:

$$p(x, y, y, t | x_0, y_0, z_0) \tag{7}$$
$$= \frac{1}{L} \sum_{l,m,n=0}^{\infty} \exp\left(-\pi^2 \cdot Dt \cdot \left(\frac{l^2}{L^2} + \frac{m^2}{L^2} + \frac{n^2}{L^2}\right)\right)$$
$$\cdot \cos\left(l \cdot \pi \cdot \frac{x - R_m}{L}\right) \cdot \cos\left(l \cdot \pi \cdot \frac{x_0 - R_m}{L}\right)$$
$$\cdot \cos\left(m \cdot \pi \cdot \frac{y - R_m}{L}\right) \cdot \cos\left(m \cdot \pi \cdot \frac{y_0 - R_m}{L}\right)$$
$$\cdot \cos\left(n \cdot \pi \cdot \frac{z - R_m}{L}\right) \cdot \cos\left(n \cdot \pi \cdot \frac{z_0 - R_m}{L}\right)$$

where $x_0, y_0, z_0$ are the initial coordinates of the Brownian particle in the box, $L$ is the size of the box, and $R_m$ is the half length of the box. It is assumed that the box is centered on the origin with fixed length $L$ on all axes.

Previously, algorithms have been proposed to sample new positions in the case of Brownian diffusion with reflective boundaries by sampling a Green function like that seen above. A possible approach would be to sample the distribution using the rejection sampling technique. Here we propose a simple but efficient algorithm which does not need to go through the rejection-acceptance method.

In Cartesian coordinates, the displacement along one axis is independent of the displacement along the other axes. Let us consider the general case of a 1D Brownian diffusion constrained by $R_m \leq x \leq R_M$ and $L = R_M - R_m$. Knowing the initial coordinate $x_0$, we sample an unconstrained position:

$$x = \text{Gauss}(0, \sqrt{2Dt}) + x_0$$

where $\text{Gauss}(x, y)$ samples a Gaussian distribution of mean $x$ and standard deviation $y$.

We compute the new position with:

$$x_{new} = R_m + h \cdot L + (1 - 2 \cdot h) \cdot |(x - R_m) \bmod L|$$

and:

$$h = \text{trunc}\left(\frac{|x - R_m|}{L}\right) \bmod 2$$

where $\text{trunc}(x)$ is the truncation operator, and h can take a value of 0 or 1.

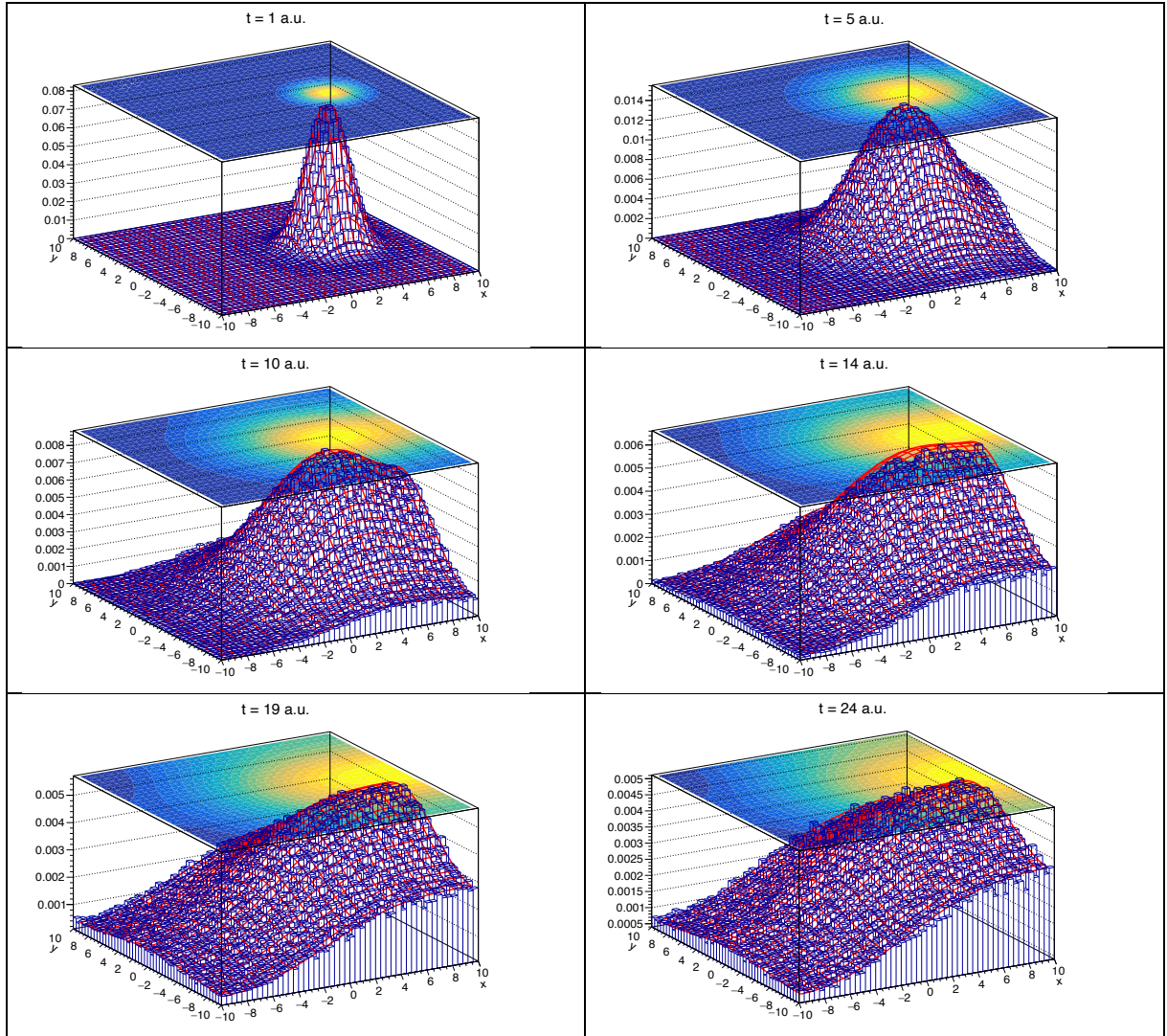

Figure 3 – Time evolution of the density probability function for a 2D diffusion constrained in a rectangle (arbitrary units). The blue histogram corresponds to the simulation results, while the red curve corresponds to the associated analytical distributions.

Figure 3 presents, for multiple time steps after initialisation, a comparison between our implementation of our algorithm against the analytical solution (equation (7)) of the reflective box model restrained to the 2D case. One million Brownian particles are initially placed at the position (4, 0, 0). In this simulation we used arbitrary units (noted as a.u.) to represent time and space, the diffusion coefficient was set to 1. Inspection of these comparative distributions illustrates a high level of agreement between the two with the time evolution of the simulation. We performed a Pearson's chi-squared test with degrees of freedom (DoF) equal to n – 1 where n is the number of histogram bins, in our case $DoF = 2499$. We hypothesize that our algorithm samples the expected distribution. We consider this hypothesize valid when $P(\chi^2 < \chi_c^2) = 95\%$ where the critical value is $\chi_c^2 \sim 2615$. We computed a $\chi^2 \sim 27.8$ which is below the chosen critical value, we can therefore not reject our hypothesize.

## 2. Analytical expressions for AA and AB reactions

### a. AA reactions

In the well stirred case for AA reactions, the analytical solution of the time evolution of the species number is given by:

$$\frac{d[A]}{dt} = -k_{obs}(t)[A]^2$$

$$\frac{1}{[A](t)} - \frac{1}{[A](0)} = \int_0^t k_{obs}(s) \cdot ds$$

$$[A](t) = \frac{1}{\int_0^t k_{obs}(s) \cdot ds + \frac{1}{[A](0)}}$$

$$N_A(t) = \frac{1}{\frac{\int_0^t k_{obs}(s) \cdot ds}{V} + \frac{1}{N_A(0)}} \tag{8}$$

where $k_{obs}(t)$ can either be the time-dependent reaction rate constant or the ultimate constant $k_{obs}(\infty)$, depending on the time scale of interest (frequency at which the reactions occur).

### b. AB reactions

The analytical solution for the AB reaction can be expressed in terms of the ratio of the numbers of the two species involved:

$$\frac{d[A]}{dt} = \frac{d[B]}{dt} = -k_{obs}(t) \cdot [A][B]$$

$$\frac{N_B(t)}{N_A(t)} = \frac{N_B(0)}{N_A(0)} \cdot \exp\left([N_B(0) - N_A(0)] \cdot \frac{\int_0^t k_{obs}(s) \cdot ds}{V}\right). \tag{9}$$

## 3. Benchmarking

As our IRT implementation was developed for unconstrained diffusion, a number of minor adjustments were required for fair comparison to the analytical reflective box model. These adjustments were made to correct for reflections at the container edge, whilst preserving the fundamental event-driven nature of the IRT method (i.e. the simulation steps from one reactant pair interaction event to another). We do not have specific diffusion events triggered to account for the reflection on the edges. This approach for the diffusion algorithm was implemented to prevent the simulation from stopping whenever a species reaches an edge of the box and to increase computational efficiency. However, in contrast to this, the IRT method must constrain the time steps taken to an upper boundary to ensure that the reflection is correctly taken into account when sampling reactions. Therefore, in the case of the reflective box model, we trigger the diffusion events only if all computed reaction times are greater than a set threshold, or after a reaction occurs and only if products are accounted.

Table 1 – Set of species and diffusion coefficients

| Species | Diffusion coefficient ($10^{-9}\ m^2 \cdot s^{-1}$) |
|---|---|
| $e_{aq}^-$ | 4.5 |
| $H_3O^+$ | 9 |
| $\cdot H$ | 7.9 |
| $\cdot OH$ | 2.8 |

In our comparison, we set the maximum time step to be:

$$dt_{max} = \frac{h^2}{6D}$$

where $h$ is the half size of the box and $D$ is the largest diffusion coefficient involved in the simulation. A reaction event having a time step larger than this limit is disregarded and a diffusion step is applied instead. The lower limit of the simulation time steps was set to be:

$$dt \gg \frac{mD}{k_B T}$$

where $m$ is the reduced mass of the Brownian particles. This is a necessary condition of the Smoluchowski diffusion model validity. If the lower limit of reaction time is not respected, it would result in the diffusing species traveling at velocities approaching infinite values. Therefore, when reaction times are reported to be smaller than this lower limit, the reaction is immediately calculated in a null time step, resulting in no diffusion of any the species. Fortunately, such small time steps are reported only in rare cases and mimic the behavior of a dense environment.

Table 2 – Set of studied reactions

| Reactions | $k_{obs}$ ($10^{10}\ L \cdot mol^{-1} \cdot s^{-1}$) | Reference | $k_{act}$ ($10^{10}\ L \cdot mol^{-1} \cdot s^{-1}$) | Reference |
|---|---|---|---|---|
| $e_{aq}^- + H_3O^+$ | 2.11 | 10 | 2.53 | 10 |
| $\cdot OH + \cdot OH$ | 0.454 | 25 | 1.03 | 26 |
| $e_{aq}^- + e_{aq}^-$ | 0.5 | 25 | / | / |
| $e_{aq}^- + \cdot H$ | 2.5 | 25 | / | / |

We compared the results of our IRT implementation to that obtained by the analytical model for a sufficiently large number of chemical species distributed homogenously throughout a 100 × 100 × 100 μm reflective box of liquid water. The set of chemical species that were investigated and their corresponding diffusion coefficients are summarized in Table 1. Whereas the studied reactions for these chemical species can be found in Table 2. Each simulation was initialized to the conditions outlined in Table 3, and over a time domain of 0.1 to 10,000 seconds the concentration of each species was tracked. As can be seen in Figure 4, under these conditions, an excellent agreement is observed between our implemented version of IRT and the analytical model.

Table 3 – Reaction types, initial conditions and models used

|  | Reaction | Initial number | Model used | Type |
|---|---|---|---|---|
| AA reactions | $\cdot OH + \cdot OH$ | 1000 | Eq. (1) | Partially diffusion-controlled reaction between neutral species |
| | $e_{aq}^- + e_{aq}^-$ | 1000 | Eq. (3) | Diffusion-controlled reaction between ionic species |
| AB reactions | $e_{aq}^- + \cdot H$ | $H = 700$<br>$e_{aq} = 500$ | Eq. (2) | Diffusion-controlled reaction between neutral species |
| | $e_{aq}^- + H_3O^+$ | $H_3O^+ = 1000$<br>$e_{aq} = 500$ | Eq. (4) | Partially diffusion-controlled reaction between ionic species |

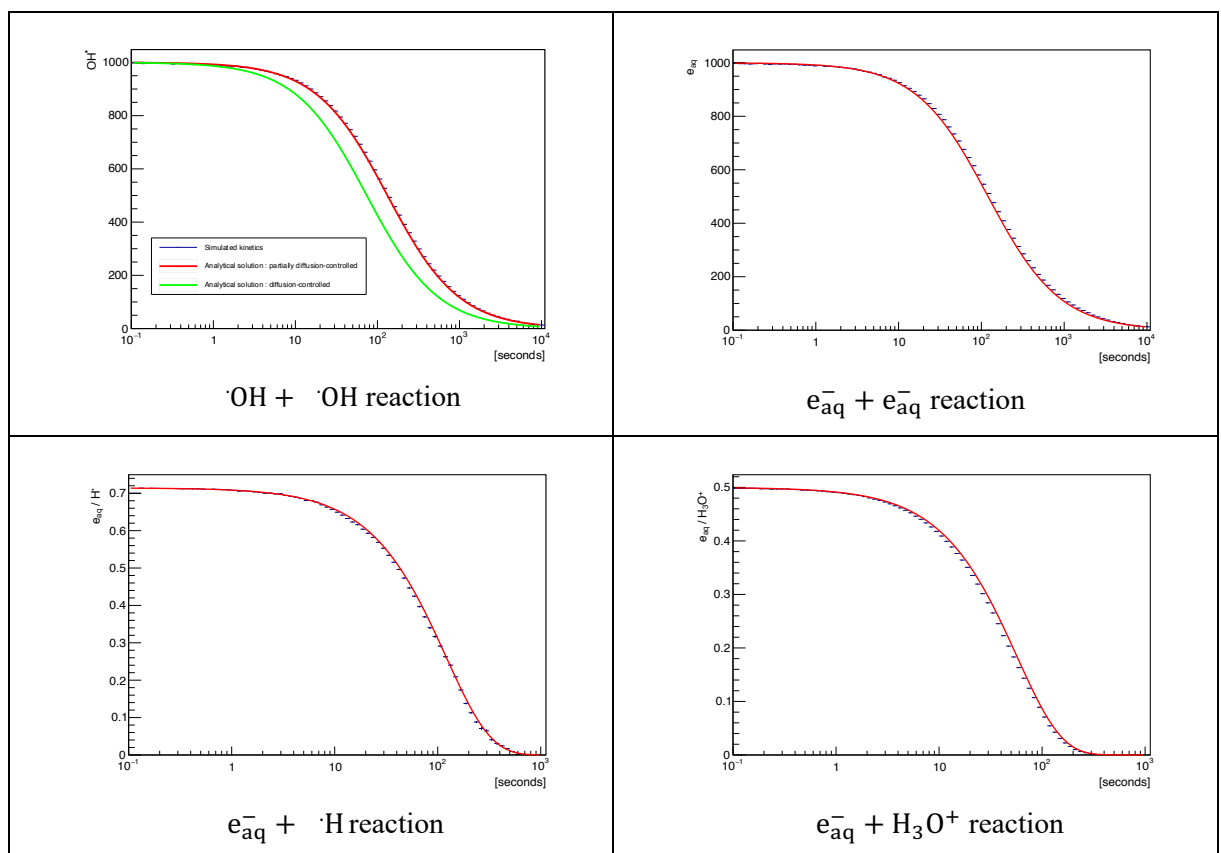

Figure 4 – Comparison between IRT (blue histogram) and the analytical expression (red and green curves) in a 100 µm side box

## Conclusions

We implemented the independent reaction times method together with (partially) diffusion-controlled pair survival models for neutral and ionic species in the Geant4 toolkit. In order to test our implementation of the method and the associated pair survival models, we restricted ourselves to study the dynamics of well stirred bimolecular systems confined in a fully reflective box. To ensure a fair comparison, we implemented an algorithm to implicitly account for the bouncing of the species on the walls of the reflective box in the diffusion engine of the IRT method. We rederived the reaction rates of the relevant bimolecular reactions, used these to benchmark the implemented models, and verified that the implemented IRT method faithfully reproduces the analytical solutions.

## Acknowledgments

J.A.L. was supported by the Division of Chemical Sciences, Geosciences and Biosciences, Basic Energy Sciences, Office of Science, United States Department of Energy through grant number DE-FC02-04ER15533. This contribution is NDRL-5238 from the Notre Dame Radiation Laboratory.